\documentclass[a4paper]{jpconf}

\usepackage{graphicx}
\usepackage{epsfig}
\usepackage{comment}
\usepackage{amssymb}
\usepackage{footmisc}
\usepackage{amsmath}
\usepackage{hyperref}

\makeatletter
\newcommand{\distas}[1]{\mathbin{\overset{#1}{\kern\z@\to}}}%
\newsavebox{\mybox}\newsavebox{\mysim}
\newcommand{\distras}[1]{%
  \savebox{\mybox}{\hbox{\kern3pt$\scriptstyle#1$\kern3pt}}%
  \savebox{\mysim}{\hbox{$\to$}}%
  \mathbin{\overset{#1}{\kern\z@\resizebox{\wd\mybox}{\ht\mysim}{$\to$}}}%
}
\makeatother

\begin{document}

\title{Low Energy Nuclear Structure from Ultrarelativistic Heavy-Light
  Ion collisions~\footnote{Talk by ERA at 37th Brazilian Workshop on
    Nuclear Physics, 8-12 September 2014, Maresias, SP, Brazil}}

\author{Enrique Ruiz Arriola$^1$
 and Wojciech Broniowski$^{2,3}$}

\address{$^1$ Departamento de F\'{\i}sica At\'{o}mica, Molecular y Nuclear and
Instituto Carlos I de  F{\'\i}sica Te\'orica y Computacional, 
Universidad de Granada, E-18071 Granada, Spain}

\address{$^2$ Institute of Physics, Jan Kochanowski University, 25-406~Kielce, 
Poland}

\address{$^3$ The H. Niewodnicza\'nski Institute of Nuclear Physics PAN, 31-342
Cracow, Poland}

\ead{earriola@ugr.es (ERA)}
\ead{Wojciech.Broniowski@ifj.edu.pl (WB)}

\begin{abstract}
The search for specific signals in ultrarelativistic heavy-light ion
collisions addressing intrinsic geometric features of nuclei may open
a new window to low energy nuclear structure.  We discuss specifically
the phenomenon of $\alpha$-clustering in $^{12}$C when colliding with
$^{208}$Pb at almost the speed of light.
\end{abstract}

\section{Introduction}

Even before the neutron was discovered, based on the $\alpha$ decay
explained successfully as a quantum tunneling effect across the
Coulomb barrier, Gamow proposed that alpha particles ($^4$He nuclei) are
the constituents of atomic
nuclei~\cite{gamow1931constitution}. Indeed, alpha clusters are
present in light nuclei (such as, e.g., $^{12}$C) as effective degrees of
freedom and lead to large intrinsic deformation of their nuclear
distributions, which in some cases lead to polyhedral symmetric
structures:  an equilateral triangle for $^{12}$C, tetrahedron for
$^{16}$O, trigonal bipyramid for $^{20}$Ne, octahedron pentagonal bipyramid
for $^{24}$Mg, hexagonal bipyramid for 
$^{28}$Si, 
etc.~\cite{wefelmeier1937geometrisches,PhysRev.54.681,PhysRev.52.1083, 
brink1965alpha,Brink1970143},
(for reviews see, e.g.,
\cite{freer2007clustered,ikeda2010clusters,beck2012clusters,%
  Okolowicz:2012kv,Zarubin}).
\begin{figure}[h]
\begin{center}
\includegraphics[width=15pc]{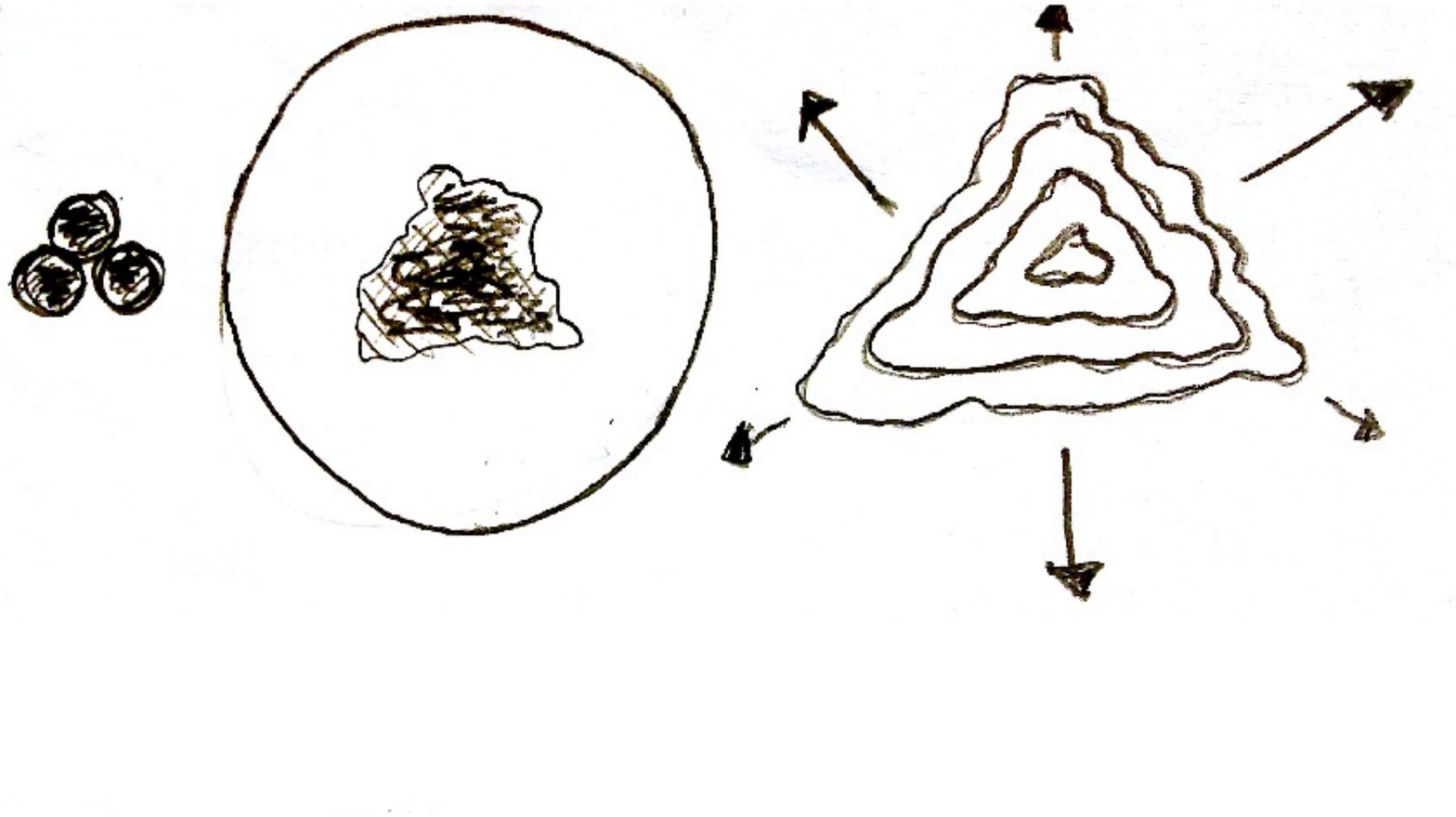}\hspace{1pc}%
\begin{minipage}[b]{17pc}\caption{ \small\label{fig:triang-wall} 
\small 
A triangle hitting the wall: $^{12}$C pictured
  as a triangle of three $\alpha$ particles (left) colliding with a heavy ion
  drawn as a round flat object (middle). The triangular shape of the 
fireball yields a
  triangular pattern of particle emission after the collision (right).
}
\end{minipage}
\end{center}
\end{figure}

When such a clustered geometric object, say $^{12}$C, hits a large
nucleus such as $^{208}$Pb at almost the speed of light (see
Fig.~\ref{fig:triang-wall}), the shape of the created fireball in the
transverse plane reflects (up to random fluctuations) the deformation
of the light nucleus and effectively implements the almost
instantaneous collapse of the collective wave
function~\cite{Broniowski:2013dia}. As the hit nucleus is large, the
fireball is abundant enough to evolve collectively, much as in
collisions of two heavy nuclei at RHIC or the LHC. Because of the
initial deformation, harmonic flow develops, leading to very specific
and measurable signatures in the transverse momentum distributions of
the detected hadrons produced in the collision. For instance, the
$^{12}$C+$^{208}$Pb system develops large triangular flow, increasing
with the multiplicity of the number the produced
hadrons. Traditionally, it has been assessed that only some very bulk
information on the nuclear structure content could be probed by
ultrarelativistic heavy-ion collisions. Our calculations show,
however, that certain very distinct geometric features, such as the
$\alpha$-clustering in the nuclear ground state wave function, could
indeed be traced significantly by properly analyzing the multiparticle
spectrum in the final state~\cite{Broniowski:2013dia,Bozek:2014cva}.

This talk returns to this remarkable link between the lowest
energy nuclear structure and the highest-energy heavy-light-ions
collisions, emphasizing some important aspects of this connection,
mostly from the nuclear physics point of view.  This is intended for
specialists from either the $\alpha-$cluster and the relativistic
heavy-ion communities who might be unfamiliar with the ideas of the
other side.

In order to motivate the presentation below, let us review the basic
elements needed to describe the collisions between two nuclei, A+B.  Let the 
two nuclei A and B be characterized by wave functions satisfying $H_A 
\Psi_{n,A}= E_{n,A}
\Psi_{n,A} $ and $ H_B \Psi_{n,A}= E_{n,A} \Psi_{n,A} $. The collision
at invariant CM energy $\sqrt{s}$, in their initial
ground state $\Psi_{0,A}$ and $\Psi_{0,B}$, can be described by the
total Hamiltonian,
\begin{eqnarray}
H= H_A + H_B + V_{AB},
\end{eqnarray}
where additivity of the elementary energy-dependent NN interactions is assumed,
\begin{eqnarray}
V_{AB} (s; \vec x_1, \dots, \vec x_A; \vec y_1, \dots, \vec y_B)
= \sum_{i=1}^A \sum_{j=1}^B V_{NN} (s,\vec x_i - \vec y_j) ,
\end{eqnarray}
with $V_{NN} (s,\vec x)= {\rm Re}V_{NN} (s,\vec x) + i {\rm Im} V_{NN}
(s,\vec x) $ denoting an optical potential. The basic tool of our analysis is
given by the Glauber formalism (see, e.g.,
\cite{Florkowski:2010zz,Florkowski:2014yza}), which gives the density
$P_{AB}(s,\vec b)$ of inelastic collisions in the transverse plane,
$\vec x = (\vec x_T, z)$, for the A+B reaction in terms of the
elementary NN density, $P_{NN}(s,\vec b)$. For instance, in the binary collision
model (the popular wounded nucleon model applied in our studies includes the 
same ingredients)
\begin{eqnarray}
P_{AB} (s,\vec b) &=& \prod_{i=1}^A \int d^3 x_i \prod_{j=1}^B \int d^3 y_j
|\psi_{0,A} (\vec x_1, \dots , \vec x_A )|^2 |\psi_{0,B} (\vec y_1, \dots ,
\vec y_B )|^2 \nonumber \\ &\times& \frac{1}{\sigma_{AB}^{\rm inel}(s)}
\left\{1- \prod_{i=1}^A \prod_{j=1}^B \left[1- \sigma_{NN}^{\rm inel} (s)
  P_{NN}(s,\vec b- \vec x_{i,T}+ \vec y_{j,T}) \right] \right\},
\label{eq:glauber-AB}
\end{eqnarray} 
where $\sigma_{NN}^{\rm inel} (s)$ and $\sigma_{AB}^{\rm inel} (s)$ are the
inelastic cross sections for NN and AB, respectively, and $\int d^ 2 b
P_{NN}(s,b)= \int d^ 2 b P_{AB}(s,b)=1 $. 
This formula shows that we need a knowledge of the nuclear wave
functions and the interaction between nucleons in the different
nuclei. This is a purely hadronic picture of the collision process and
the only partonic underlying features are hidden under $P_{NN}(s,\vec b)$
which is usually obtained phenomenologically. 

\section{The $\alpha$-particle}

\subsection{Basic properties}

\begin{figure}[t]
\begin{center}
\epsfig{figure=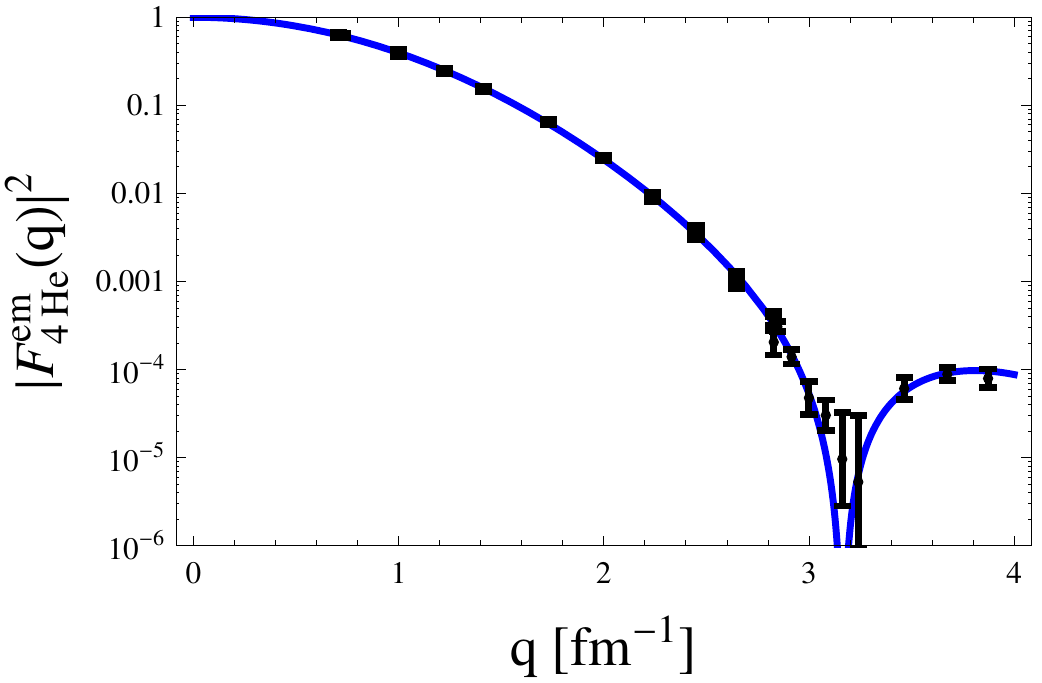,width=7.5cm}\hskip1cm 
\epsfig{figure=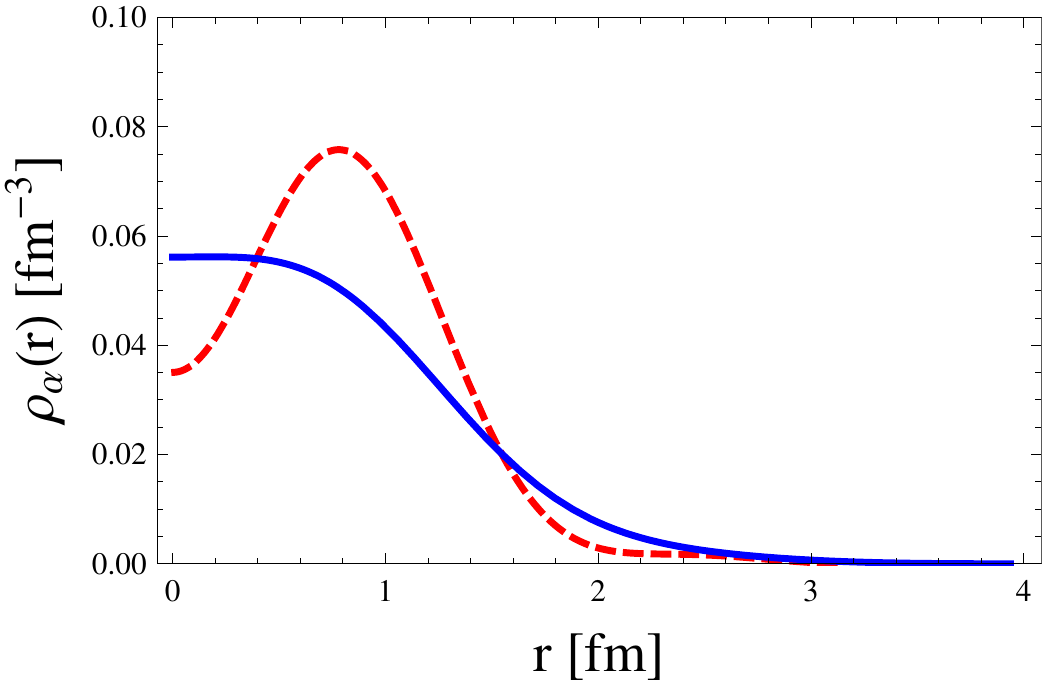,width=7.5cm}
\end{center}
\vspace{-5mm}
\caption{ \small Left: Charge form factor for $^4$He as a function of
  the momentum transfer (data from Ref.~\cite{frosch1967structure}).
  Right: Normalized charge (full line) and baryon (dashed line)
  density for the $\alpha$ particle as functions of the radial
  distance.}
\label{fig:ff-He4}
\end{figure}

The $\alpha$ particle is a $^4{\rm He}=ppnn$ nucleus, which in the
ground state has the quantum numbers $(J^P, T)=(0^+,0)$.  Its 
charge is $Z_\alpha e = + 2 e $ and mass $M_\alpha = 2(M_p+M_n) -
B = 3727.37~{\rm MeV}$ with the binding energy $B=28.2957~{\rm
  MeV}$. There are no excited states below the $t+p$ or $^3$He+n
thresholds, and the nucleon separation energy is $S_n \sim 20~{\rm
  MeV}$. The size information is obtained  from the electron
scattering, where the  elastic $e + ^4{\rm He} $  differential cross section
is given by
\begin{eqnarray}
\left(\frac{d \sigma}{d \Omega} \right)_{e,\alpha}= \frac{d \sigma_{\rm Ruth}}{d
  \Omega} |F_{^4{\rm He}}^{\rm em} (q)|^2,
\end{eqnarray}
deviating from the point-like Rutherford result. The charge form
factor can be well fitted by~\cite{frosch1967structure}
$F_{^4{\rm He}}^{\rm em} (q)= \left(1 - q^{12} c^{12} \right)
e^{-a_{em,\alpha}^2 q^2} $, with $ a_{em,\alpha}=0.681(2)~{\rm fm} $
and $ c=0.316(1)~{\rm fm}$, and vanishes at $q \simeq 3~{\rm fm}$, see left
Fig.~\ref{fig:ff-He4}.  The corresponding charge mean squared radius is $\langle
r^2 \rangle_{em,\alpha} = 6 a_{em, \alpha}^ 2= (1.668 (5)~{\rm
  fm})^2$. The nucleon distribution is obtained by unfolding the
charge distribution with the nucleon (proton) charge form factor
$F^{\rm em}_p(q)=1/(1+q^2/\Lambda^2)^2$, $\Lambda^2=0.5 {\rm GeV}^2$:
\begin{eqnarray}
F^{\rm em}_\alpha (q)= \int d^3 x e^{i \vec q \cdot \vec x} \rho^{\rm em}_\alpha (r) = 
Z_\alpha F^{\rm em}_p (q)  \int d^3 x e^{i \vec q \cdot \vec x} \rho_\alpha (r)  \, \, 
\label{eq:ff-em-alpha}
\end{eqnarray} 
and assuming isospin invariance. Then the baryon density $\rho_B (r) = 4
\rho_{\alpha} (r)$. Normalized charge and baryon densities are depicted in
right Fig.~\ref{fig:ff-He4}.  Thus, we have $\langle r^2
\rangle_\alpha = \langle r^2 \rangle_{em,\alpha} - \langle r^2
\rangle_{em,p}= (1.42~{\rm fm})^2$.

\subsection{Simple  wave functions}
A good and simple approximation for the wave function in the CM
system, as far as the form factor is concerned for $c=0$ in
Eq.~(\ref{eq:ff-em-alpha}), is given by the shell model as a $ |(1s)^4
\rangle $ state,
\begin{eqnarray}
\Psi_\alpha ( \vec x_1, \vec x_2, \vec x_3, \vec x_4) &=&  \prod_{i=1}^4 \frac{e^{-\frac{(\vec x_i-\vec R)^2}{2b^2}}}{(\sqrt{\pi} b)^{3/2}}  {\cal A} (n\uparrow, n\downarrow, p\uparrow, p\downarrow ) \, , 
\end{eqnarray}
where ${\cal A}$ is a normalized antisymmetrizer and $\vec
R=\sum_{i=1}^4 \vec x_i/4$ is the CM. The ground state probability
$|\Psi_\alpha (\vec x_1, \vec x_2, \vec x_3 , \vec x_4)|^2$ can be
constructed by randomly generating 12 Gaussian variables describing
$\vec x_i$ of the four particles, and then relocating the CM $\vec R =
\sum_{i=1}^4 x_i/4 $ to the origin. More realistic wave functions can
be achieved by including a short-distance correlation features, say a
hard core of size $d$, by excluding configurations with $|\vec x_i -
\vec x_j| \le d$~\cite{Broniowski:2010jd}. This simple approach will
be pursued below for the $^{12}$C ground state wave function.  Further
improvements incorporate asymptotic long-distance cluster
decomposition conditions on the total wave function into $^2$H+$^2$H,
$^3$H+n or $^3$He+p subsystems~\cite{Schiavilla:1985gb}, implying
exponential fall-offs with proper separation wave numbers.

\section{$\alpha$-$\alpha$ interaction and effective elementarity} 

The study of the $\alpha$-$\alpha$ interaction, involving eight
nucleons, has been a continuous playground in cluster model
studies~\cite{Tang:1978zz,Friedrich:1981ad}. The total wave function
is written as $\Psi_{\alpha,\alpha} = \Psi_{\alpha} (1,2,3,4)
\Psi_{\alpha} (5,6,7,8) \chi_{\alpha \alpha} (r) $, where the relative
wave function satisfies a RGM equation containing a direct term which
is {\it local} and corresponds to a folding
structure~\cite{Satchler:1979ni} very much like the nucleus-nucleus
Sao Paulo potential~\cite{Chamon:2002mx}, and an exchange term which
is non-local and short range.

\subsection{$\alpha$-$\alpha$ potential}

\begin{figure}[ttt]
\begin{minipage}{18pc}
\includegraphics[width=18pc]{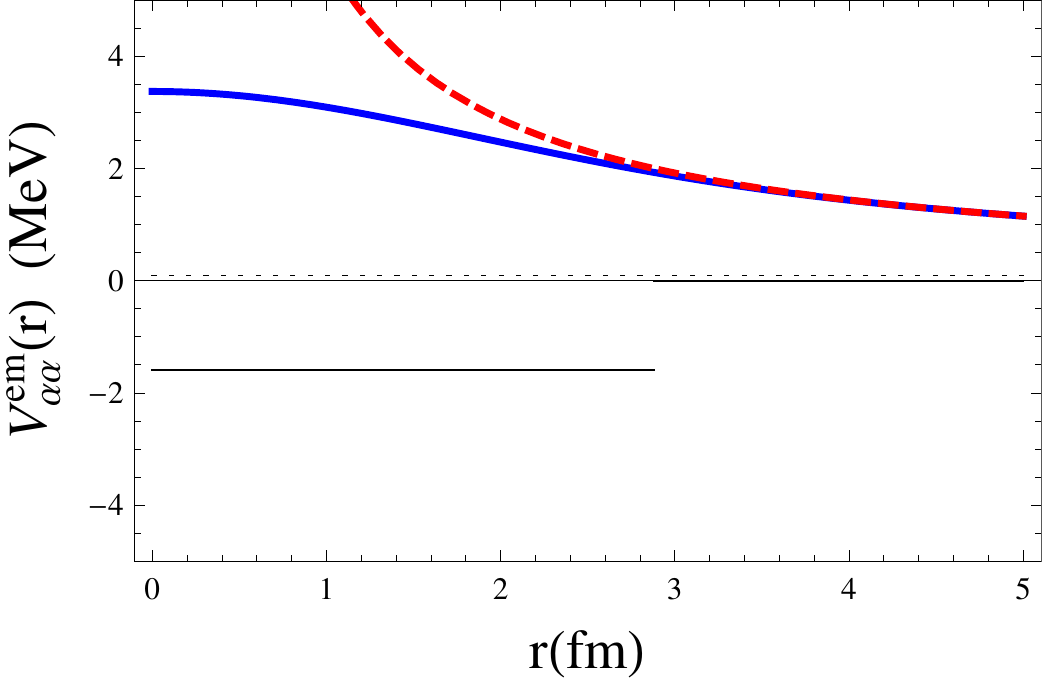}
\caption{ \small\label{Vaa-em} Direct $\alpha$-$\alpha$
  electromagnetic potential (in MeV) as a function of the distance (in
  fm).  We compare the folding potential (solid line) with the
  elementary Coulomb potential (dashed line), the $Q=92
  \,{\rm KeV}$ value for $^8$Be (dotted line) and the
  $B_{\alpha\alpha}=1.52 {\rm MeV}$ binding when Coulomb is removed
  above $r_c = 2.88 \,{\rm fm}$ (solid broken line).}
\end{minipage}\hspace{2pc}%
\begin{minipage}{18pc}
\includegraphics[width=18pc]{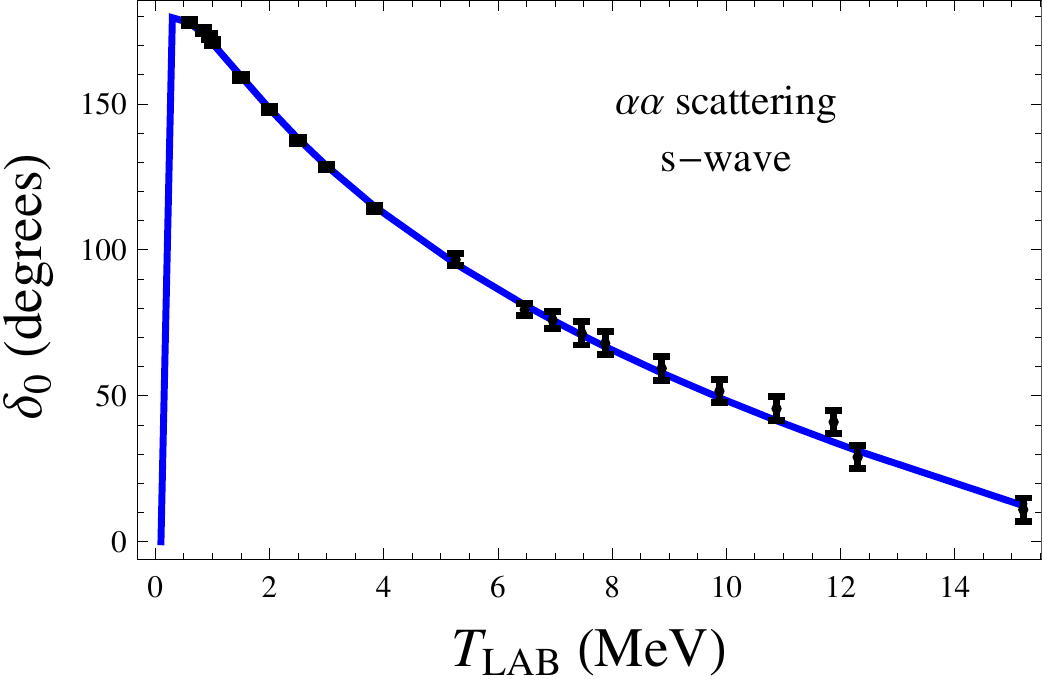}
\caption{ \small\label{ps-alpha-alpha} S-wave phase shift for the
  $\alpha$-$\alpha$ scattering as a function of the LAB energy.  The
  line is the fit using a boundary condition at a distance of
  $r_c=2.88 {\rm fm}$ and elementary Coulomb potential above
  $r_c$. The data can be traced from Ref.~\cite{AFZAL:1969zz}. }
\end{minipage} 
\end{figure}

The direct electromagnetic interaction between the two $\alpha$ particles
can be written as a folding potential, leading in the overlapping
region to the screening of the pure Coulomb interaction below an
elementarity radius, $r_c$,
\begin{eqnarray}
V_{\alpha\alpha}^{\rm em}(r) = \int d^3 \vec r_1 d^3 \vec r_2 \frac{\rho_\alpha^{\rm em}(\vec r_1)\rho_\alpha^{\rm em} (\vec r_2)}{|\vec r_1-\vec r_2 - \vec r|} 
= \int \frac{d^3  \vec q}{(2\pi)^3} \frac{4 \pi e^ 2}{\vec q^2} |F_{\alpha}^{\rm em}(\vec q)|^2 e^{i \vec q \cdot \vec r} 
\sim \frac{Z_\alpha^2 e^ 2}{r}, \qquad r \gtrsim r_c. \nonumber 
\end{eqnarray}
The leading exchange interaction corresponds to separating one neutron
or proton, leaving $^3$H or $^3$He behind, respectively, and is ${\cal
  O} (e^{- 2 \gamma_{n,t} r})$, where $S_n = \gamma_{n,t}^2/(2 \mu_{n,t})$
the nucleon separation energy. Using the nuclear form factor, see
Fig.~\ref{fig:ff-He4}, we immediately see from Fig.~\ref{Vaa-em} that
$\alpha$ particles interact effectively as if they were point-like for
distances larger than $r \gtrsim 2.5~{\rm fm}$. We call this {\it
  effective elementarity}, a feature that entitles us to treat
interactions between $\alpha$ particles as point-like objects above
this separation~\footnote{Due to the quantum numbers of the
  $\alpha$ particle, the longest range non-em interaction is of van der
  Waals nature and is given by Two-Pion-Exchange, a tiny
  effect~\cite{Arriola:2007de,RuizArriola:2008cy}. Note that this
  notion of elementarity has to do with the interaction; the simple
  overlap integral $\int d^3 s \rho_\alpha (s-r) \rho_\alpha (s)
  \approx 0$ for $r=r_c$ is not the relevant property.} and disregard
the short-range exchange term. One interpretation of the
$\alpha$ clustering is based on the fact that in $A=4n$ nuclei
$\alpha$ particles are on the average packed at a larger distance than
$r_c$, and thus preserve their identity. 

\subsection{$\alpha$-$\alpha$ scattering}

Assuming this simple picture, we can analyze $\alpha$-$\alpha$ scattering
without much detailed knowledge of the internal $\alpha$ particle
structure at sufficiently low energies, with an upper bound  fixed by the
relative CM momentum $p= \hbar /r_c \sim 200~{\rm MeV}$ corresponding
to $T_{LAB} = 2p^2 /M_\alpha\sim 15~{\rm MeV}$. The scattering
amplitude is written as a partial wave expansion~\cite{AFZAL:1969zz}
\begin{eqnarray}
f(\theta)= f_C (\theta)+ \sum_{l=0}^\infty (2l+1)e^{2 i
  \sigma_l(p)}\frac{e^{2 i \delta_l(p)}-1}{2 i p}
P_l(\cos\theta), \label{eq:famp}
\end{eqnarray}
where $f_C (\theta)$ is the Coulomb amplitude, $\sigma_l(p)$ and
$\delta_l(p)$ are the Coulomb and strong phase-shifts, respectively,
obtained from solving the relative $\alpha$-$\alpha$ wave function
$\chi_{\alpha,\alpha} (r) \equiv u_{l,p}(r)/r$ with the boundary
condition $ u_{l,p} (r) \sim G_l(\eta, p r) \sin \delta_l (p)+ F_l
(\eta, p r ) \cos \delta_l (p)$. Here $G_l (\eta, p r)$ and $F_l
(\eta, p r )$ are Coulomb wave functions and $\eta= 1/(pa_B) $ is the
Sommerfeld parameter and $a_B = 2/(M_\alpha Z_\alpha^2 e^2)= 3.63\, {\rm fm}$ the Bohr radius. Due to the Bose statistics, we have $l=0,2,4,
\dots$.  The partial wave expansion, Eq.~(\ref{eq:famp}), is limited
in practice to $l_{\rm max} \sim p a$, with $a$ denoting the range of
the non-electromagnetic interaction, since classically $L = p b $ and
there is no scattering for $b> a$. Below the first inelastic $^7$Li+p
and $^7$B+n thresholds, which take place at $T_{\rm LAB} \sim 15 {\rm
  MeV}$, we have $l_{\rm max} \sim 4$. Using an energy independent
boundary condition at a given cut-off radius $r_c$ we get for $l=0$
(S-wave)
\begin{eqnarray}
- \frac1{M_\alpha}u_{0,p}'' (r)+ \frac{e^2 Z_\alpha^2}{r} u_{0,p} (r) =
  \frac{p^2}{M_\alpha} u_{0,p} (r), \qquad  \quad L=\frac{u_{0,p}' (r_c)}{u_{0,p} (r_c)},
\end{eqnarray}
from which we can fit the S-wave phase-shift with the
parameters~\cite{Arriola:2007de,RuizArriola:2008cy}
\begin{eqnarray}
L= -0.357(3)~{\rm fm}^{-1}, \qquad r_c = 2.88(3)~{\rm fm}, \qquad \chi^2
/{\rm DOF} = 0.5. \, , 
\end{eqnarray}
which is rather good, see Fig.~\ref{ps-alpha-alpha}.  Thus, all the
necessary information is encoded in these parameters, in harmony with
an effective elementarity of $\alpha$ particles and the irrelevance of
the short distance characteristics. Therefore, as long as $\alpha$
particles remain at distances larger than $\sim 3~{\rm fm}$, we may
ignore all the substructure details, including the exchange terms.
That way one can sidestep a full {\it ab initio} calculation with 8
nucleons.

\subsection{$\alpha$-$\alpha$ binding and tunneling}

The steep rise of the phase-shift starting at zero corresponds to the
$^8$Be resonance, which arises as a pole of the S matrix in the second
Riemann sheet. The asymptotics $ u_R (r) \to e^{i p r} $ with the
complex momentum $ p= p_R + {\rm i} p_I $ gives a complex energy $ E =
Q - i\Gamma /2 = (p_R+ {\rm i} p_I)^2/(2 \mu_{\alpha\alpha}) = 92~{\rm
  KeV} - \frac{\rm i}2 3.4(2)~{\rm eV} $, so that the turning point,
fulfilling $Q= V^{\rm em}_{\alpha\alpha} (r_0)$, is rather large $r_0=
62 {\rm fm}$, yet tunneling through the barrier occurs. A further
interesting aspect is that if the Coulomb force were removed above
$r_c=2.88~{\rm fm}$, the $^ 8$Be resonance would then become a stable
weakly bound state with a binding energy given by
$B_{\alpha\alpha}=-\gamma_B^2 /M_\alpha = -1.5229~{\rm MeV}$ (see
Fig.~\ref{Vaa-em}), a fairly large radius, $\langle r^2
\rangle^\frac12 = 4.0764~{\rm fm}$, and the asymptotic coefficient of
the normalized reduced wave function, $u_B(r) \to A_B e^{-\gamma_B r}$
of $A_B = 2.4163~{\rm fm}^{-1/2}$.

\section{Evidence for $\alpha$-clustering}

The effective elementarity of $\alpha$ particles has far reaching
consequences. The most clear signatures for $\alpha$ clustering in
nuclei are seen in the binding energies and form factors. Already in
1966 Harrington proposed to treat $^{12}C$ as a bound state of three
elementary $\alpha$ particles~\cite{PhysRev.147.685}.

\subsection{Binding energies}

\begin{figure}
\begin{center}
\epsfig{figure=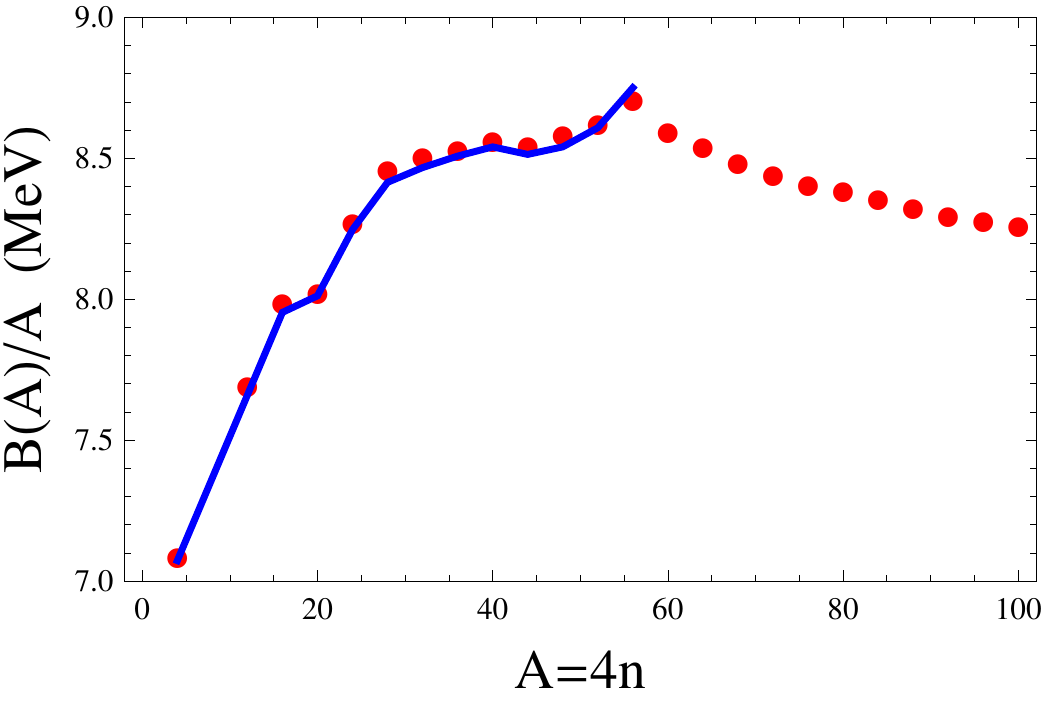,width=7.5cm}\hskip1cm
\epsfig{figure=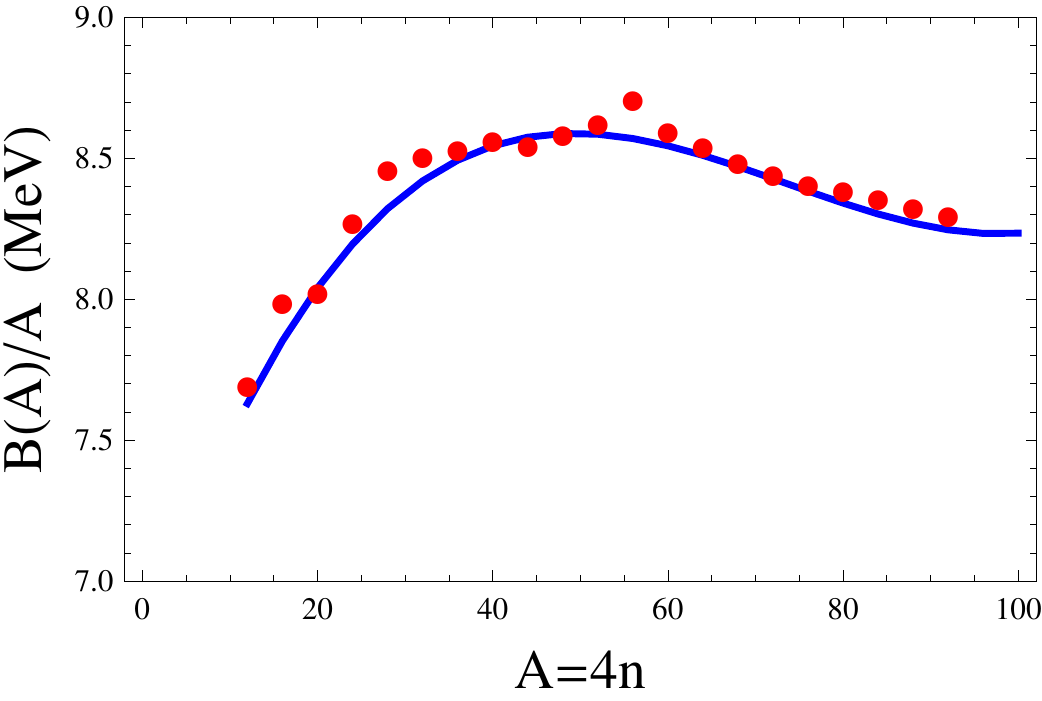,width=7.5cm}
\caption{ \small Binding energy per particle  as a function of the
  mass number for $A=4n$ nuclei. Left: Bonding geometric model. Right:
  Many-body model containing two-,three- and four-$\alpha$ forces.}
\label{fig:binding}
\end{center}
\end{figure}

Since the $\alpha$ particle is very tightly bound, a
hint for clustering for $A=4n$ nuclei was given by the bonding
approximation~\cite{PhysRev.54.681,RevModPhys.35.40},
\begin{eqnarray}
E(4n)=-B(4n) = -n B_\alpha + n_{\rm bonds} V_{\alpha \alpha} \qquad V_{\alpha \alpha} = -2.29 {\rm MeV}  \, ,  \qquad n \ge 3 \, , 
\label{eq:bonding}
\end{eqnarray}
where $n_{\rm bonds}=3,4,8,12,16,19,22,25,27,30$ for $^{12}$C,
$^{16}$O, $^{20}$Ne, $^{24}$Mg, $^{28}$Si, etc., respectively,
yielding $n_{\rm bonds}/n \to 2.5$ (we stop plotting at $A=56$ since
$n_{\rm bonds}$ can always be adjusted onwards). This corresponds a
next-neighbors interaction of closely packed spheres (see
\cite{freer2007clustered} for pictures illustrating the polyhedral
configurations)~\footnote{The case of $^{20}$Ne would require $n_{\rm
    bonds}=9$ for this interpretation but that gives a worse
  binding.}. The result of the fit can be seen in
Fig.~\ref{fig:binding} (left).
One drawback of this approach is that the fit does not reproduce the simplest
$n=2$ case, which corresponds to $^8$Be.

In a different scheme we just count the numbers of pairs, triplets and
quartets, corresponding to two-,three- and four body forces, and the energy reads
\begin{eqnarray}
E(4n) = -n B_\alpha + 
\left( \begin{array}{c} n \\ 2
     \end{array} 
\right) V_{2 \alpha}+
\left( \begin{array}{c} n \\ 3
     \end{array} 
\right)
V_{3 \alpha}+
\left( \begin{array}{c} n \\ 4
     \end{array} 
\right) V_{4 \alpha} \, ,    \qquad n \ge 3 \, , 
\label{eq:234-alpha}
\end{eqnarray} 
A fit for $A=4n$ nuclei gives $ V_{2\alpha}= -2.39~{\rm MeV}, \,
V_{3\alpha}= 0.47~{\rm MeV}, \, V_{4\alpha}= -0.04~{\rm MeV} \, $. The
result can be seen in Fig.~\ref{fig:binding} (right), with a mean
standard deviation of $0.1~{\rm MeV}$ per nucleon. Again, the meaning
of $n=2$ is particularly subtle since, as we have shown, this
corresponds to the $^8$Be case which is unbound by $\sim 0.092~{\rm
  MeV}$ while direct extrapolation of the general formula for $n$
implies $ \sim -2.4~{\rm MeV}$. The $^8$Be is an exceptionally large
system which becomes very sensitive to the long range Coulomb
interaction. As we have seen elimination of $V_{\alpha\alpha}^{\rm em}
(r)$ above $r_c$ gives $B_{\alpha \alpha} \sim 1.52~{\rm MeV}$, see
Fig.~\ref{Vaa-em}, reducing partly the puzzle in the $n=2$ case both
for the bonding model, Eq.~(\ref{eq:bonding}), and the many body
model, Eq.~(\ref{eq:234-alpha}).

\subsection{Cluster Sizes}

\begin{figure}[ttt]
\begin{center}
\epsfig{figure=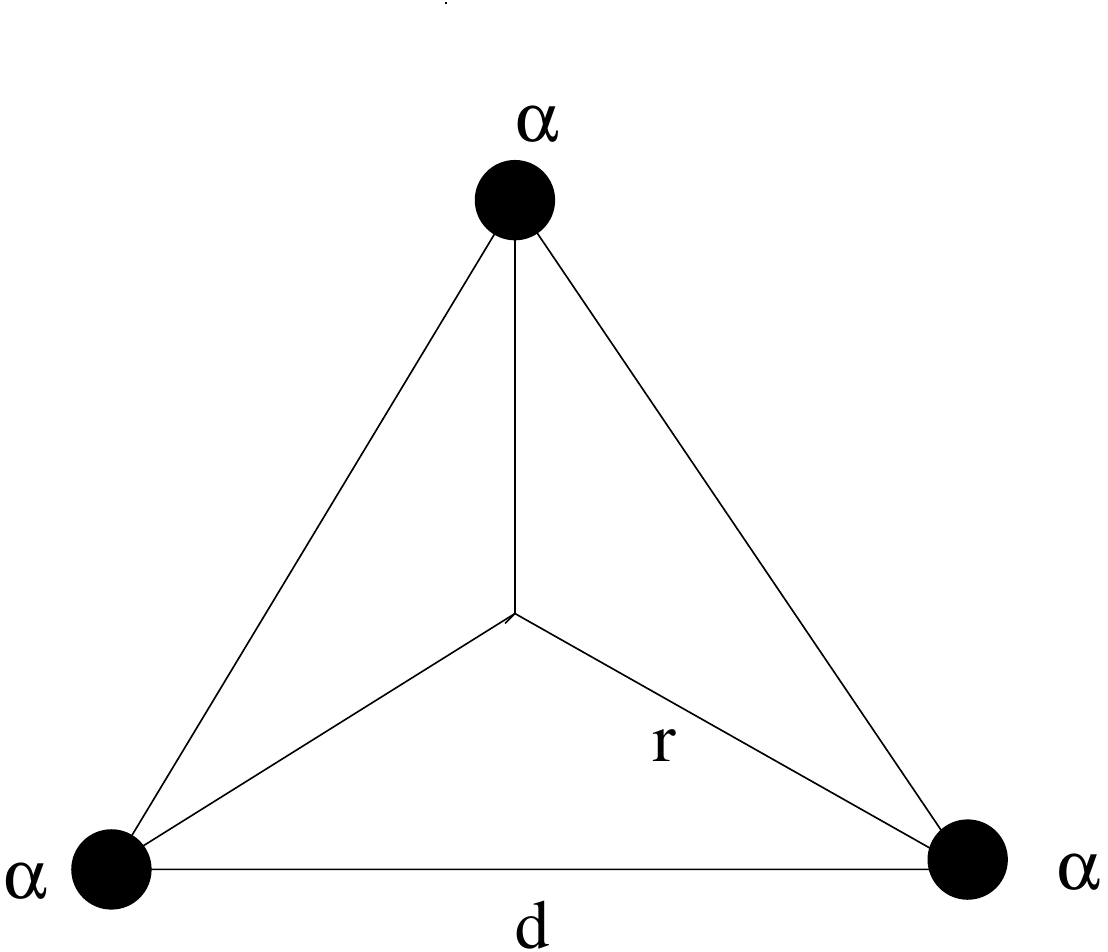,width=5cm}\hskip0.5cm
\epsfig{figure=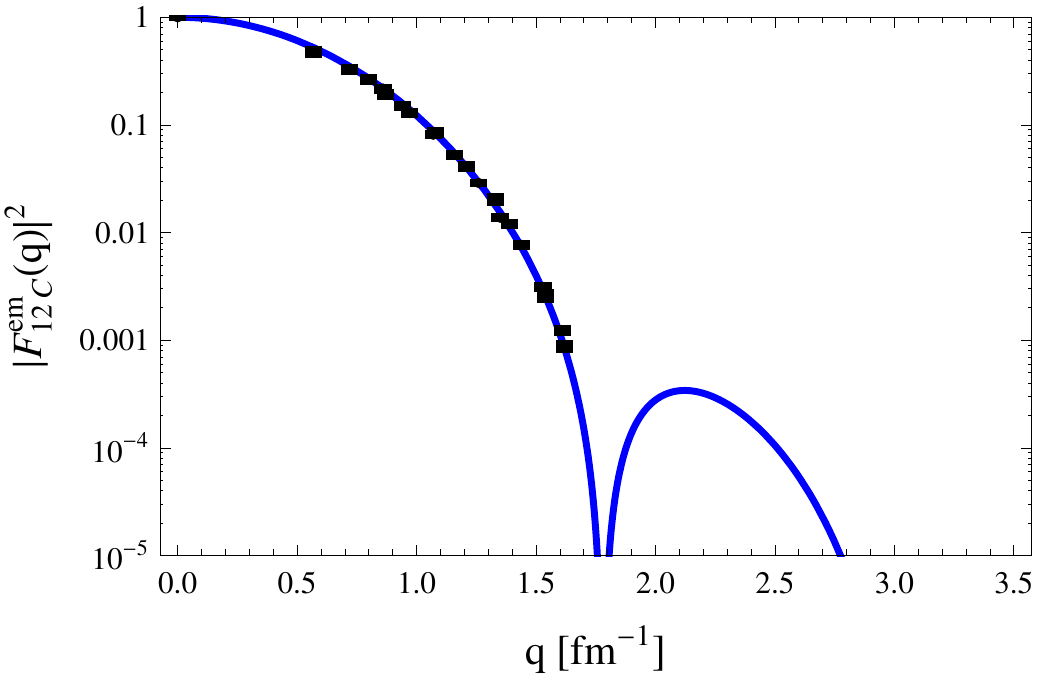,width=8cm}
\end{center}
\caption{ \small Left: Triangle structure in $^{12}$C.  Right: Corresponding
  charge corm factor of $^{12}$C from folding the $\alpha$ charge form
  factor with the triangle $F_{^{12}C}^{\rm em}(q) = F_{^4He}^{\rm em}(q)
  F_{\vartriangle}(q)$ with $d=3 {\rm fm}$ (data from
  Ref.~\cite{sick1970elastic}).}
\label{fig:ff-12C}
\end{figure}

Charge form factors provide valuable insight on cluster geometries and
sizes.  If we focus on $^{12}C$, we fold an equilateral triangle with
side $d$ and radius $r_\vartriangle = d/\sqrt{3}$ with an
$\alpha$ particle, see Fig.~\ref{fig:ff-12C},
\begin{eqnarray}
F_{^{12}C}^{\rm em}(q) = F_{^4He}^{\rm em}(q)
  F_{\vartriangle}(q) \, , 
\qquad F_{\vartriangle}(q) = 
\frac{\sin (q d/\sqrt{3})}{q d /\sqrt{3}} \, ,
\end{eqnarray}
A fit to the experimental elastic form factor of
$^{12}$C~\cite{sick1970elastic} taking the experimental $F_{^4He}^{\rm
  em}(q)$ (see Eq.~\ref{eq:ff-em-alpha} and Fig.~\ref{fig:ff-He4})
yields $d = 3.05~{\rm fm} > r_c=2.88~{\rm fm}$, consistent with the
effective elementarity.

\subsection{Modeling clustering}

Besides this very simple but direct evidence, a massive effort is
being carried
out~\cite{freer2007clustered,ikeda2010clusters,beck2012clusters,Okolowicz:2012kv})
to probe the clusterization features of light nuclei.  Clustering in
$^{12}$C has been described by the Bose-Einstein Condensation (BEC)
model~\cite{Funaki:2006gt}, the fermionic molecular dynamics
(FMD)~\cite{Chernykh:2007zz}, antisymmetrized molecular
dynamics~\cite{KanadaEn'yo:2006ze}, effective chiral field theory on
the lattice~\cite{Epelbaum:2012qn}, the no-core shell
model~\cite{Roth:2011vt,Barrett:2013nh}, or the variational
Green's function method (VMC)~\cite{Pieper:2002ne}.  The recently
discovered $5^-$ rotational state of $^{12}$C in low energy
$\alpha+^{12}C$ collisions points to the triangular ${\cal D}_{3h}$
symmetry of the system~\cite{Marin-Lambarri:2014zxa}. While it would be best to
incorporate realistic calculations (see,
e.g.,~\cite{Buendia:2004yt,Wiringa:2013ala}), in
Ref.~\cite{Broniowski:2013dia} we have applied a simple and practical
procedure with $\alpha$ clustered (or unclustered for comparison)
random distributions constructed as follows.  In the clustered case we
randomly generate positions of the 12 nucleons, 4 in each cluster of a
Gaussian shape and size $r_\alpha$. The centers of the clusters are
placed in an equilateral triangle of side length $d$, cf. 
Fig.~\ref{fig:ff-12C}.  The
distribution of the 12 nucleons is recentered such that the center of mass
is placed at the origin.  The short-distance NN repulsion is incorporated by
precluding the centers of each pair of nucleons to be closer than the
expulsion distance of 0.9~fm~\cite{Broniowski:2010jd}.

\begin{figure}[tb]
\begin{center}
\epsfig{file=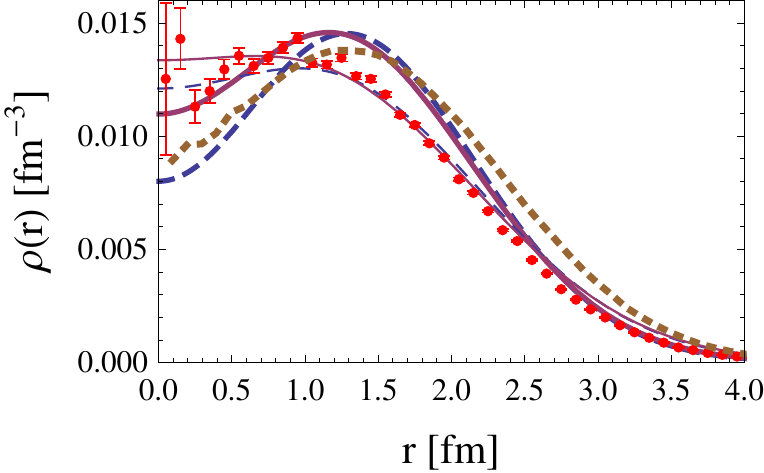,width=7.5cm,height=5cm}
\epsfig{file=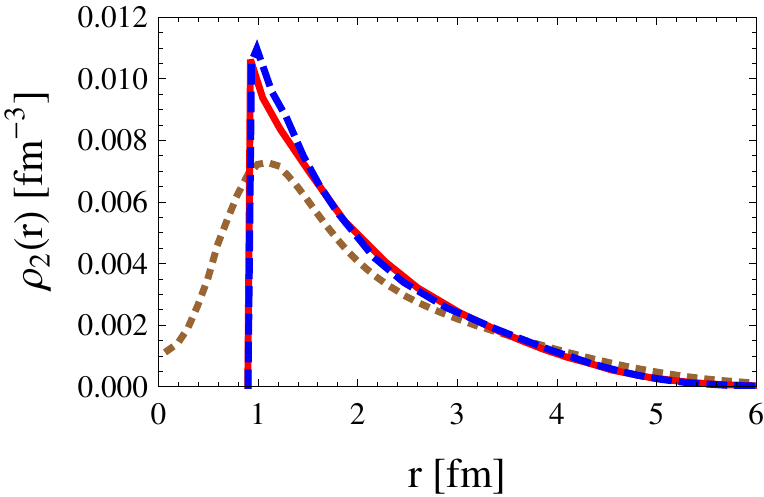,width=7.5cm,height=5.2cm}
\end{center}
\caption{ \small\label{fig:dense} \small Left: Normalized one-particle
distributions in $^{12}C$.
The electric charge density $\rho^{\rm em}(r)/Z$ (thin lines) and the
corresponding distribution of the centers of nucleons $\rho (r)$
(thick lines) in ${}^{12}$C for the data~\cite{DeJager:1987qc} and
BEC~\cite{Funaki:2006gt} calculations (dashed lines),
FMD~\cite{Chernykh:2007zz} calculations (solid lines), and Jastrow
correlated wave function~\cite{Buendia:2004yt} (dotted line).  
Right: Normalized two particle distribution
$\rho_2(r)$ in $^{12}C$. We show our results for the fitted $\rho(r)$
in the FMD (solid line) and BEC cases (dashed line), and
compare to the Jastrow correlated wave function~\cite{Buendia:2004yt}
(dotted line).} 
\end{figure}

The single and pair distribution functions are defined as (see
~\cite{Viollier:1976ab} for multiple distributions)
\begin{eqnarray}
\rho (\vec x) &=&\langle \Psi_A | \hat \rho(\vec x) | \Psi_A \rangle 
 = \int d^3 x_2 \dots d^3 x_A |\Psi_A(\vec x , \vec x_2, \dots, \vec x_A)|^2, 
\nonumber \\ \rho_2 (\vec r) &=& 
 \int d^3 R
d^3 x_3 \dots d^3 x_A |\Psi_A(\vec R + \frac12 \vec r, \vec R - \frac12 \vec r , \vec x_3,  
\dots, \vec x_A)|^2,
\end{eqnarray}
and are normalized to unity, i.e. $\int d^3 r \rho (\vec r)=
\int d^3 r \rho_2 (\vec r)=1$, admitting a simple probability interpetation. 
These are the densities corresponding to point-like nucleons, hence the
charge form factor reads,
\begin{eqnarray}
F_A^{\rm em} (q) = Z F_p^{\rm em} (q) \int d^3 x \rho (\vec x) e^ {i \vec x \cdot \vec q},
\end{eqnarray}
where $F_p^{\rm em} (q)$ is the proton charge form factor (we neglect
the small neutron charge form factor and assume isospin
symmetry). These densities have exponential fall-offs,
\begin{eqnarray}
\Psi_A (\vec x , \vec x_2 , \dots, \vec x_A) & \distras{x\to \infty} & Z_A 
\chi_1 (\vec x) \Psi_{A-1} (\vec x_2 , \dots, \vec x_A), \nonumber \\
\Psi_A (\vec x , \vec x', \vec x_3, \dots, \vec x_A) & \distras{x,x'\to \infty} 
& Z_A \chi_{12} (\vec x, \vec x') \Psi_{A-2} (\vec x_3 , \dots, \vec x_A),
\end{eqnarray}
because of the cluster decomposition. In
Refs.~\cite{Schiavilla:1985gb,Wiringa:1991kp} the calculations
explicitly implemented this exact asymptotic property, which would
become relevant when separation energies are small or in the case of
halo nuclei, which is not the case for $^{12}C$ in the ground state. 

The parameters $d$ and $r_\alpha$ are optimized such that the one
particle density $\rho(r)$ of the BEC~\cite{Funaki:2006gt} or
FMD~\cite{Chernykh:2007zz} calculations are accurately reproduced (as
explained, the unfolding of the proton charge density from the charge
distribution $\rho^{\rm em}(r)$ is necessary), see
Fig.~\ref{fig:dense}.  Note a large central depletion in the
distributions, originating from the separation of the $\alpha$
clusters. Besides, a fair reproduction of the two-particle densities,
$\rho_2(r)$, obtained from multiclustered Jastrow correlated
calculations~\cite{Buendia:2004yt} is observed.  The radial
distribution $4\pi r^2\rho_2(r)$ peaks at the size of the triangle, $d
\simeq 3~{\rm fm}$~\cite{Broniowski:2014aqa}. Thus, in our simulations
we deal with simplified but realistic nuclear distributions of
$^{12}$C.

\section{Ultrarelativistic collisions}

We now sketch the basic ingredients behind Eq.~(\ref{eq:glauber-AB}).
We start by analyzing the NN collisions at high energies, where the
Mandelstam $s= 4 (p^2 + M_N^2)$ in the CM frame.

\begin{figure}[ttt]
\begin{center}
\epsfig{figure=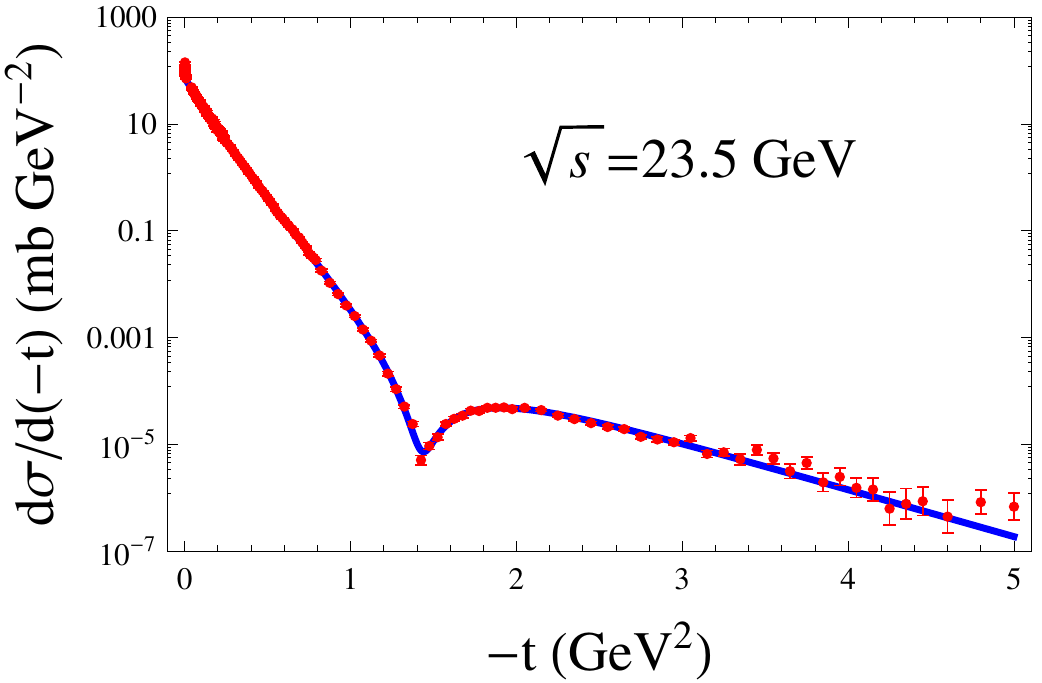,width=7.5cm}\hskip.3cm
\epsfig{figure=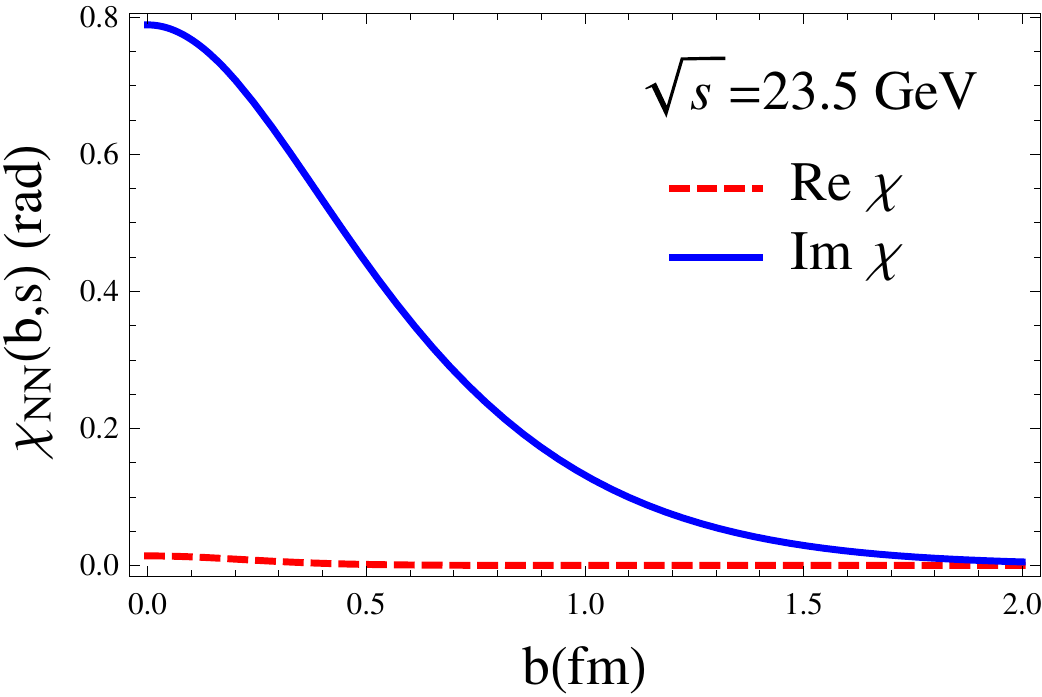,width=7.5cm}\vskip.3cm
\epsfig{figure=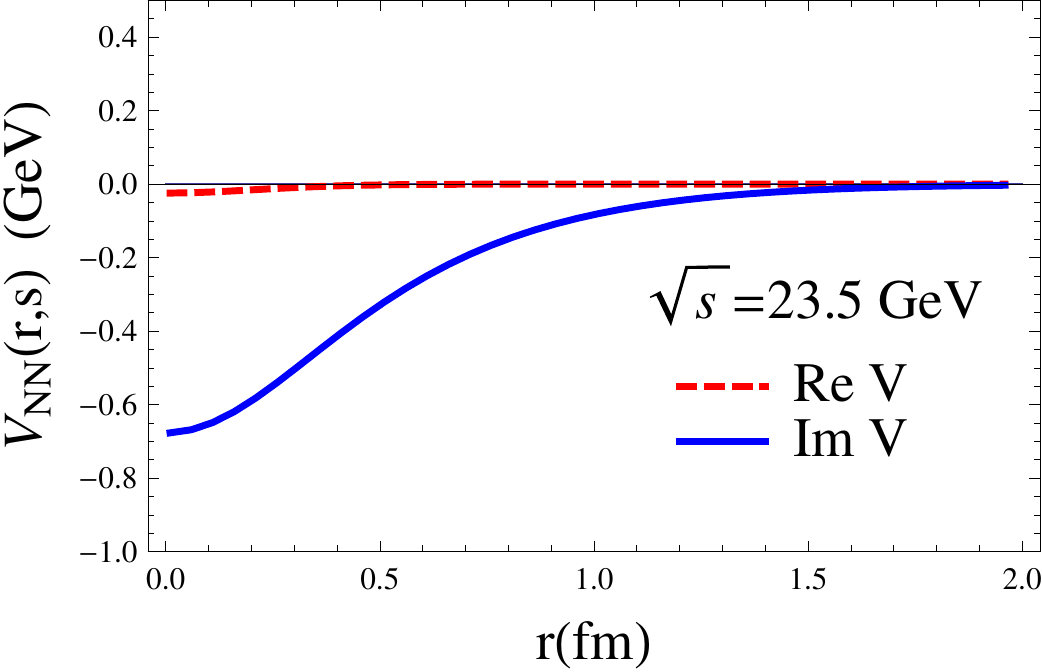,width=7.5cm}
\end{center}
\caption{ \small NN scattering at CM energy $\sqrt{s}=23~{\rm
    GeV}$. Top Left: Differential cross section as a function of the
  invariant squared momentum transfer $-t$ vs. data
  from~\cite{Amaldi:1979kd,Amos:1985wx}. Top Right: Eikonal real and
  imaginary phase as a function of the impact parameter. Bottom:
  Optical potential as a function of the distance.}
\label{fig:NN-23}
\end{figure}

\subsection{NN collisions at high energies} 

Neglecting spin effects, the scattering amplitude can be written as in
Eq.~(\ref{eq:famp}) and in the limit $l_{\rm max} \sim p a \gg 1 $ we
can replace the sum over $l$ by an integral.  Using the impact parameter
as the integration variable $b \, p = l+ 1/2$, the formula $P_l (\cos \theta) 
\to
J_0 (bq) $, and dropping the Coulomb 
interaction, yields~\cite{Florkowski:2010zz}
\begin{eqnarray}
f_{NN \to NN}(s,t)= \frac{i p }{2\pi}\int d^2 b [1-e^{2 i \delta(b)}]
e^{i \vec b \cdot \vec q}.
\end{eqnarray}
At high energies relativistic effects and inelasticities must be taken
into account. The general field theoretic approach would require a
coupled channel Bethe-Salpeter equation, where the kernel would
ultimately be determined phenomenologically from the NN scattering
data. Under these circumstances, it is far simpler to use a minimal
relativistic approach based on the squared mass
operator~\cite{Allen:2000xy}.  Here $ {\cal M}^2 = P^\mu P_\mu + W $,
where $W$ represents the (invariant) interaction determined in
the CM frame by matching to the non-relativistic limit with an energy-dependent 
and local optical potential, $V(\vec r,s)= {\rm Re } V(\vec
r,s) + i {\rm Im} V(\vec r,s)$. This yields for NN after quantization  $ {\cal \hat M}^2 =
4(\hat p^2+M_N^2) + 8 M_N V $, with $\hat p = - i \nabla$, 
so that the relativistic equation can be
written as ${\cal \hat M}^2 \Psi = 4 (k^2+M_N^2) \Psi $, i.e.,  as a
non-relativistic Schr\"odinger equation
\begin{eqnarray}
(-\nabla^2 + M_N V ) \Psi = (s/4-M_N^2) \Psi \, . 
\end{eqnarray}
In the eikonal approximation the (complex) phase-shift is given by
\begin{eqnarray}
\delta_l(k)\Big|_{l+\frac12= bk} \approx
- \frac{2\mu}{2k}\int_b^\infty dr \frac{r V(r,s)}{\sqrt{r^2-b^2}} \equiv \frac12 
\chi(b,s) =  \frac12 {\rm Re}\chi(b,s) + i \frac12 {\rm Im}\chi(b,s),
\end{eqnarray}
where $\chi(b,s)$ is the eikonal phase.  The optical potential
$V(r,s)$ can be obtained from a fit to the experimental elastic
differential cross section $d\sigma/d(-t) = |f|^2 \pi/p^2$. The fits
for pp scattering at $\sqrt{s}=23.5~{\rm
  GeV}$~\cite{Amaldi:1979kd,Amos:1985wx} are depicted in
Fig.~\ref{fig:NN-23} for illustration (details and further results
will be presented elsewhere). As we see, both the phase shift and the optical
potential are almost purely imaginary, i.e., ${\rm Re} \chi(b) \ll {\rm
  Im} \chi(b) $. This is a general feature above $\sqrt{s} \gtrsim 10~
{\rm GeV}$, which allows us to write the inelastic cross section from
the optical theorem $\sigma_{\rm tot}= 4 \pi {\rm Im} f(0) /p \sim
\sigma_{\rm inel} $,
\begin{eqnarray}
\sigma_{\rm inel} (s) = \int d^2 b \left[ 1- e^{-{\rm Im} \chi_{NN}(s,b)} 
\right] \equiv \sigma_{\rm inel} (s)  \int d^2 b P_{NN} (s,b),
\end{eqnarray}
in terms of a profile function $P_{NN}(s,b)$ interpreted
as the probability of the NN inelastic collision at
impact parameter $b$.


\subsection{Nucleus-nucleus scattering (Glauber)}

Nucleus-nucleus interactions is most conveniently studied by the
Glauber theory applied to multiple scattering (for a review see
e.g.~\cite{Miller:2007ri}.), where the highly asymmetric role played
by the $NN$ interactions within a given nucleus and $NN$ interactions
between nucleons of different nuclei is explicitly exploited. 
Actually, the interaction time is so short that only high energy
states are excited, so basically both incident nuclei remain in their ground
states. Under these assumptions, the initial wave function can be written as
$\Psi= e^{i p z } \Phi_{AB} \psi_A \psi_B$ for a collision along the
$z-$axis, with a slowly varying relative wave function $ \Phi_{AB}
$. The elastic scattering amplitude becomes
\begin{eqnarray}
f_{AB \to AB}(s,t)= \frac{i p }{2\pi}\int d^2 b 
\langle \Psi_A \Psi_B| [1-e^{i \chi_{AB}(b)}] | \Psi_A \Psi_B \rangle
e^{i \vec b \cdot \vec q}.
\end{eqnarray}
The key simplifying aspect is that from the
additivity of the NN interactions one obtains the additivity of the eikonal
phases,
\begin{eqnarray}
\chi_{AB}(b; \vec x_{1,T}
, \dots , \vec x_{A,T}; \vec y_{1,T}, \dots , \vec y_{B,T} ) = \sum_{i=1}^A 
\sum_{j=1}^B \chi_{NN} (\vec b -\vec x_{i,T} + \vec y_{j,T}).
\end{eqnarray}
A purely imaginary NN phase, $\chi_{NN}(b) \sim i {\rm Im}
\chi_{NN}(b)$, implies a purely imaginary AB-phase, $\chi_{AB}(b) \sim
i {\rm Im} \chi_{AB}(b)$, and Eq.~(\ref{eq:glauber-AB}) follows from  
using again the optical theorem for the AB case.  

The evaluation of the Glauber multidimensional integral in
Eq.~(\ref{eq:glauber-AB}), or for the applied wounded nucleon model,
can be done through the Monte Carlo techniques such as, e.g., in
GLISSANDO~\cite{Broniowski:2007nz,Rybczynski:2013yba}. Within such an
approach the two-dimensional density of sources can be constructed.
We note that a realistic distribution that can be used for the
hydrodynamic evolution requires smearing, as to make it sufficiently
smooth.  We refer to Ref.~~\cite{Broniowski:2013dia} for illustrations
resembling our cartoon of Fig.~\ref{fig:triang-wall}.

Valuable information on the collision geometry and its fluctuations
can be obtained from the study of eccentricity parameters (heavily
used in relativistic heavy-ion analyses). One introduces the Fourier
decomposition of the source density in the transverse plane {\em in a
  given event},
\begin{eqnarray}
\epsilon_n e^{i n \Phi_n} = \frac{\int d^ 2 b b^n P (b)e^{i n
    \phi}}{\int d^ 2 b b^nP(b)},
\end{eqnarray}
with the eccentricity parameters $\epsilon_n$ encoding the information on 
the shape of the fireball distribution. Here $n$ denotes the 
rank and
$\Phi_n$ is the principal axis angle. Then $\epsilon_2$ is called {\it
  ellipticity}, $\epsilon_3$ {\it triangularity}, etc. In general,
there are two sources for the $\epsilon_n$ parameters: the intrinsic mean
deformation of the fireball and fluctuations.  


\subsection{Harmonic flow}

The harmonic flow consequences, particularly elliptic and triangular
flow have been analyzed in detail in Ref.~\cite{Bozek:2014cva,Bozek:2014cya}, 
where it is shown
that the intrinsic deformation results in very characteristic behavior
of the ratios of the so called cumulant moments as functions of the number of
participant nucleons, both for the elliptic and triangular
deformations. Such ratios are insensitive to the details of 
dynamics as long as the response of the evolution is linear in the initial 
deformation, which is the fact for the studied systems~\cite{Bzdak:2013rya}.

\begin{figure}[tb]
\begin{center}
\includegraphics[width=21pc]{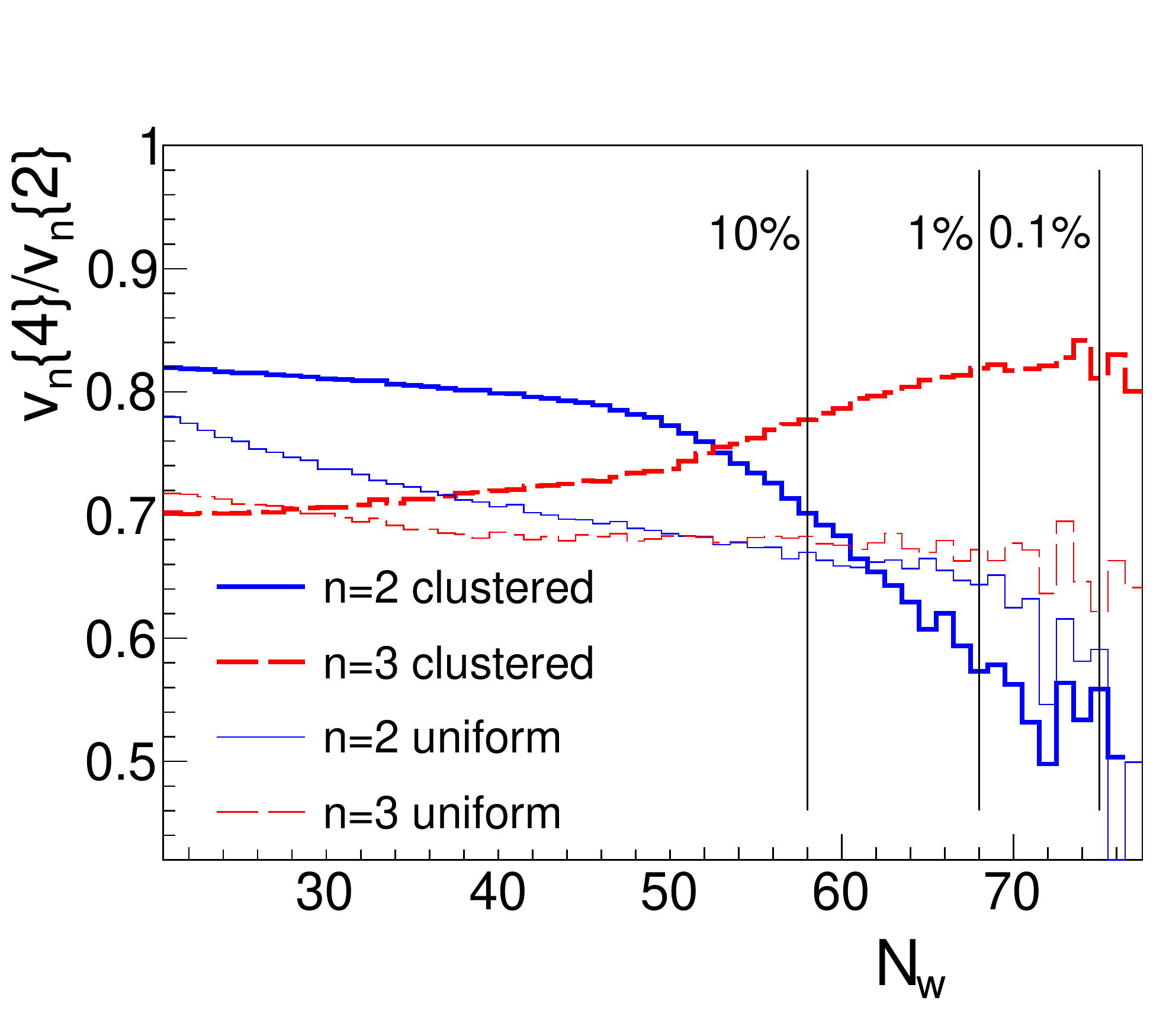}\hspace{1pc}%
\begin{minipage}[b]{15pc}\caption{ \small\label{fig:cumul} 
\small Ratios of harmonic flow coefficients of 4- and 2-particle
cumulant moments for the elliptic ($n=2$) and triangular ($n=3$) case,
plotted as functions of the number of wounded nucleons, and displaying
the radically different behavior for the clustered and uniform
$^{12}$C wave functions (thick vs. thin lines). BEC case, mixed
Glauber model for $^{12}$C+$^{208}$Pb collisions with the SPS
parameters~\cite{Bozek:2014cva}. The vertical lines indicate
centralities.}
\end{minipage}
\end{center}
\end{figure}

In Fig.~\ref{fig:cumul} we show the ratio of the four- to two-particle 
cumulant moments for ellipticity and triangularity  for the
$^{12}$C+$^{208}$Pb collisions at the highest the SPS energy of 
$\sqrt{s_{NN}}=17$~GeV. We note a significant difference in the behavior for 
the clustered wave functions of $^{12}$C (thick lines) and the uniform 
distributions (thin lines). Such effects should be the base for 
searches of clusterization signals in experiments. 

\section{Conclusions}

When the experimental data on relativistic light-heavy nuclear
collisions become available, our method, linking the lowest-energy
nuclear phenomena (alpha clustering) and the highest-energy nuclear
collisions (collective flow), will offer a novel way to study nuclear
correlations in the ground state. Conversely, a detailed knowledge of the
clustered light-nucleus wave functions will help to constrain the 
dynamics
of the fireball evolution models (hydrodynamics, transport) and thus gain 
information on the
fundamental properties of the hot and dense matter in the
fireball. With this in mind, our method, inherently investigating
multiparticle correlations leading to collective effects (harmonic flow), may
provide new insights. We stress that since the clusterization
phenomenon concerns multi-particle correlations, it is accessible
directly only through observables which are many-body. Thus the
typically studied one-body quantities, such as excitation spectra of
the EM form factors, by definition cannot ``prove'' clusterization in
a direct manner.  This unexpected bridge between these two disparate
fields of intense activity might enhance the mutual interest of
researchers in nuclear structure to produce realistic wave functions
which could be used to prepare the initial starting point for
the hydrodynamic description of relativistic heavy-ion collisions.

\ack

This research was supported by the Polish National Science Centre,
grants DEC-2011/01/D/ST2/00772 and DEC-2012/06/A/ST2/00390, Spanish
DGI (grant FIS2011-24149) and Junta de Andaluc\'{\i}a (grant FQM225).

\section*{References}

\bibliographystyle{iopart-num}

\providecommand{\newblock}{}

\end{document}